\documentstyle[epsfig,12pt]{article}
\newcommand{\beq}{\begin{equation}}
\newcommand{\eeq}{\end{equation}}
\newcommand{\bea}{\begin{eqnarray}}
\newcommand{\eea}{\end{eqnarray}}
\def\Dzero{D$\emptyset$}

\def\GeV{{\rm GeV}}
\def\MeV{{\rm MeV}}

\def\cM{{\cal M}}
\def\cR{{\cal R}}
\def\wh{\widehat}
\def\lapprox{\lower .7ex\hbox{$\;\stackrel{\textstyle <}{\sim}\;$}}
\def\gapprox{\lower .7ex\hbox{$\;\stackrel{\textstyle >}{\sim}\;$}}

\begin{document}
\titlepage
\begin{flushright}
{DTP/96/106}\\
{December 1996}\\
\end{flushright}
\begin{center}
\vspace*{2cm}
{\Large {\bf On the Partonometry of $V(=\gamma,W,Z)+$jet Events \\ [2mm]
at Hadron Colliders}} \\

\vspace*{1.5cm}
V.\ A.\ Khoze$^a$ and W.\ J.\ Stirling$^{a,b}$ \\

\vspace*{0.5cm}
$^a \; $ {\it Department of Physics, University of Durham,
Durham, DH1 3LE }\\

$^b \; $ {\it Department of Mathematical Sciences, University of Durham,
Durham, DH1 3LE }
\end{center}

\vspace*{4cm}
\begin{abstract}
Significant colour interference effects are expected 
 in $V(=\gamma,W,Z)+$jet production at hadron colliders.
 These directly influence the hadronic antenna patterns and
 can provide a valuable diagnostic tool for probing the
 nature of the underlying parton subprocesses. Motivated
 by these ideas, we present
quantitative predictions for the distributions of soft
particles and jets in $W+$jet production at the Tevatron $p\bar p$
collider.
\end{abstract}

\newpage

It was realised long ago [\ref{ref:DKT}--\ref{ref:emw}] 
that the overall 
structure of particle
angular distributions in multijet events in hard scattering 
processes (the event
portrait) is governed by the underlying colour dynamics at short
distances. A natural idea has been proposed 
[\ref{ref:DKT}--\ref{ref:ZEPPEN}]
 (for recent detailed
studies see Ref.~\cite{EKS}) to use this colour event portrait 
as a ``partonometer"
mapping the basic interaction short-distance process.
 The interest in this subject has  recently been boosted because of two
reasons. First, as was advocated in Ref.~\cite{EKS},
the hadronic antenna pattern can be used as a diagnostic tool to dissect
the colour structure of the exciting large-$E_T$ jet events observed by
CDF \cite{CDF} and D0 \cite{D0} at the Tevatron $p \bar p$ collider, 
as a way to distinguish between conventional QCD and
possible new physics production mechanisms. Secondly, both CDF and D0 have
started very successfully [\ref{ref:exp}--\ref{ref:D0Melanson}]
to measure the  structure of multijet events and have
demonstrated that the distinctive colour interference
effects survive the hadronization stage and are clearly visible in the
data.
 It has been known for a long time \cite{DKMT} that an especially
bright colour interference effect arises in the case of
large-$E_T$ production of  colour singlet objects, for instance, in
$V+$jet events (with $V=\gamma$, $W^\pm$ or $Z$).
The hadronic antenna patterns for
to such processes are entirely analogous to that in the celebrated
string \cite{string} or drag \cite{drag}
 effect in $e^+e^- \to q \bar q g$ events.

Recently the first (very impressive) data on $W+$jet production from 
D0 \cite{D0dpf96,D0Melanson} have become available.
The colour coherence effects
are clearly seen and in our view these studies have a very promising
future. They may play the same role for hadron colliders as
the important series of  results on inter-jet
studies at $e^+e^-$ colliders.

The purpose of this paper is to present
quantitative predictions for the colour
interference phenomena in the distribution of soft particles and jets
in $V+$jet production  at hadron colliders,
in particular the Tevatron $\sqrt{s} = 1.8$~TeV $ p \bar p$ collider.

There are two leading-order processes, $q \bar q \to Vg$ and
$qg \to Vq$. Each has its own distinctive antenna pattern, as we
shall see. In principle, the antenna pattern could be used
as a `partonometer' to identify the dominant scattering process.

There are two experimental methods which are likely to prove
useful. With sufficiently high statistics, one could use
an additional soft jet as the probe of the antenna pattern.
Alternatively, with fewer events one could use the distribution
of soft hadrons without
requiring jet reconstruction. Both of these quantities are directly
 related to the inclusive soft gluon distribution in the
NLO processes $q \bar q \to V g g $ and $q g \to V q g $.
We will consider such distributions, first in the soft-gluon
approximation where the results are particularly simple, and then
using the exact QCD matrix elements and phase space constraints.

The distribution of soft gluon radiation is controlled by
the basic antenna pattern (see for example Ref.~\cite{book})
\beq
[ij] \equiv {p_i \cdot p_j \over p_i\cdot k\; p_j \cdot k}
 = { 1 - {\bf n}_i \cdot {\bf n}_j \over
E_k^2 ( 1 - {\bf n}_k \cdot {\bf n}_i)\;
( 1 - {\bf n}_k \cdot {\bf n}_j)} \; ,
\label{antenna}
\eeq
where the $p_i^\mu = E_i(1,{\bf n}_i)$
are the four-momenta of the partons
participating in the hard scattering process.
This corresponds to the emission of soft primary gluons with 
energies $E_k$:
$\Lambda_{QCD} \ll E_k \ll E_i, E_j$.
The particle flow
is then described by extra multiplicative cascading factors, see below.
Since the parton scattering process acts as a colour antenna, the
distribution of soft particle flows between the jets is determined
by the overall colour structure of the event 
[\ref{ref:DKT}--\ref{ref:book},\ref{ref:drag}]

In order to make the study more quantitative we must  define
appropriate kinematic distributions and then compute the contributions
of the various subprocesses. In order to fully understand the
differences between these, we first consider the soft gluon
distribution at the parton-parton scattering level with fixed
kinematics. The generic process is
\begin{equation}
a(p_1) + b(p_2)\to c(p_3) + d(p_4) + g(k) \; ,
\label{eq:generic}
\end{equation}
where the  gluon is soft relative to the two 
large-$E_T$ partons $c$ and $d$. We assume massless quark 
and gluon partons
and write $M_V$ for the mass of the vector boson.
In our quantitative studies, we are particularly
interested in the angular distribution of the soft gluon around
one of the large-$E_T$ final-state particles, $p_3$ say. 
Using the notation $p^\mu = (E,p_x,p_y,p_z)$, we write
\begin{eqnarray}
p_3^\mu &=& (M_T\cosh\eta, 0, E_T, M_T\sinh\eta) \; , \nonumber \\
k^\mu &=& (k_T\cosh(\eta+\Delta\eta), k_T\sin\Delta\phi,
 k_T \cos\Delta\phi, k_T\sinh(\eta+\Delta\eta)) \; ,
\label{eq:4vecs}
\end{eqnarray}
where $M_T = E_T$ for $c=q,g$ and 
$M_T = \sqrt{E_T^2 + M_V^2}$ for $c=V$. 
The phase-space separation between the soft gluon and parton $c$ 
is parametrized by $\Delta\eta$ and $\Delta\phi$. Alternative
variables, more suited to the experimental analysis, are the radial and
polar angle variables in the   ``LEGO plot":
\begin{eqnarray}
\Delta\eta &=&   \Delta R    \cos\beta \; , \nonumber \\
\Delta\phi &=&  \Delta R    \sin\beta \; .
\end{eqnarray}
  which are defined in such a way 
that the LEGO--plot separation between soft jet 
$k$ and hard jet/vector boson $p_3$ is 
$R(k,p_3) = \sqrt{\Delta\eta^2+
\Delta\phi^2} = \Delta R$, ($0 \leq \Delta R
< \infty$), and the
azimuthal orientation of  $k$ around  $p_3$ in the LEGO plot
is parametrized by the angle $\beta$, ($0 \leq \beta < 2 \pi$). 
Our convention is such that $\beta = 0$ corresponds to the direction
of the incoming parton $a$, i.e. $p_{1z} > 0$. 

In terms of these variables, the soft gluon phase space is
\begin{equation}
{1\over (2 \pi)^3} \; {d^3 k \over  2 E_k}
 = {1 \over 16\pi^3}\; k_T d k_T \; \Delta R d \Delta R\;  d \beta\; .
\end{equation}
We will be particularly interested in the behaviour of the cross
section as a function of $\beta$ for fixed $k_T, \Delta R$
and fixed $E_T, \eta$.

The matrix elements for the lowest-order
processes (for the case $V=W$) are:\footnote{CKM factors
are omitted for clarity.}
\begin{eqnarray}
\overline{\sum}\vert{\cal M}\vert^2(q(p_1)\bar q(p_2) \to W(p_3) g(p_4))
 &= &  
 \frac{g_s^2g_W^2}{4} 
\;    \left(1-\frac{1}{N_c^2}\right)\; 
\frac{t^2+u^2+2sM_W^2}{tu}\; ,  \nonumber \\
\overline{\sum}\vert{\cal M}\vert^2(q(p_1)g(p_2) \to W(p_3) q(p_4))
 &= &
 \frac{g_s^2g_W^2}{4}
 \; \frac{1}{N_c}\;
\frac{t^2+s^2+2uM_W^2}{-ts}  \; ,
\label{eq:2to2}
\end{eqnarray}
where $s=(p_1+p_2)^2$, $t=(p_1-p_3)^2$ and $u=(p_1-p_4)^2$.
In the soft-gluon approximation, the corresponding
$2\to 3$ matrix elements are \cite{DKMT}
\begin{eqnarray}
\overline{\sum}\vert{\cal M}\vert^2(q\bar q \to W g(g))
 &= & g_s^2 N_c \left( [14] + [24] -\frac{1}{N_c^2} [12] \right)\;  
 \overline{\sum}\vert{\cal M}\vert^2(q\bar q \to W g) \nonumber \\
\overline{\sum}\vert{\cal M}\vert^2(qg \to W q(g))
 &= & g_s^2 N_c \left( [12] + [24] -\frac{1}{N_c^2} [14] \right)\;
\overline{\sum}\vert{\cal M}\vert^2(qg \to W q)
\label{eq:2to3}
\end{eqnarray}
Note that for these processes, the effect of the soft gluon emission
is simply to multiply the lowest-order matrix elements squared
by an overall factor consisting of three different antennae,
defined according to (\ref{antenna}), one of which is suppressed
in the large $N_c$ limit. This structure is universal for any
electroweak boson $+$ jet production, i.e. $V = W^\pm , Z^0, \gamma,
\gamma^*$. Although we shall only study the $V=W$ case in detail,
similar results can be obtained for the other cases 
as well.\footnote{In particular, the results for $Z^0+$jet production
will be almost identical to those for $W+$jet production, 
but in practice
the event rates are an order of magnitude smaller.}

To illustrate the colour coherence properties of the above
scattering processes, we
first show results for a very simple kinematic configuration,
$2\to 2$ scattering at $90^{\circ}$ in the parton centre-of-mass
frame, i.e. $\eta=0$ in Eq.~(\ref{eq:4vecs}). For the case when
the `target' final-state particle (with momentum $p_3$) is a $W$
boson, the matrix elements squared are as given above and 
the various
antennae which contribute to the radiation pattern are,
in terms of the variables $\Delta R$ and $\beta$,
\begin{eqnarray}
{}[12] & = & 2 \; ,                        \nonumber \\
{}[14] & = & {\exp(\Delta R\cos\beta)\over \cosh(\Delta R\cos\beta)
+ \cos(\Delta R\sin\beta) }   \; ,           \nonumber \\
{}[24] & = &   {\exp(-\Delta R\cos\beta)\over \cosh(\Delta R\cos\beta)
+ \cos(\Delta R\sin\beta) }   \; .
\label{eq9}
\end{eqnarray}
where an overall factor of $1/k_T^2$ has been omitted from the right-hand
sides. When the target final-state particle is the quark or gluon,
i.e. for the subprocesses $q\bar q \to g W$ and $q g \to q W $,
the corresponding results are obtained by interchanging the momenta
$p_3$ and $p_4$ in Eq.~(\ref{eq:2to3}).

\begin{figure}[tb]
\begin{center}
\mbox{\epsfig{figure=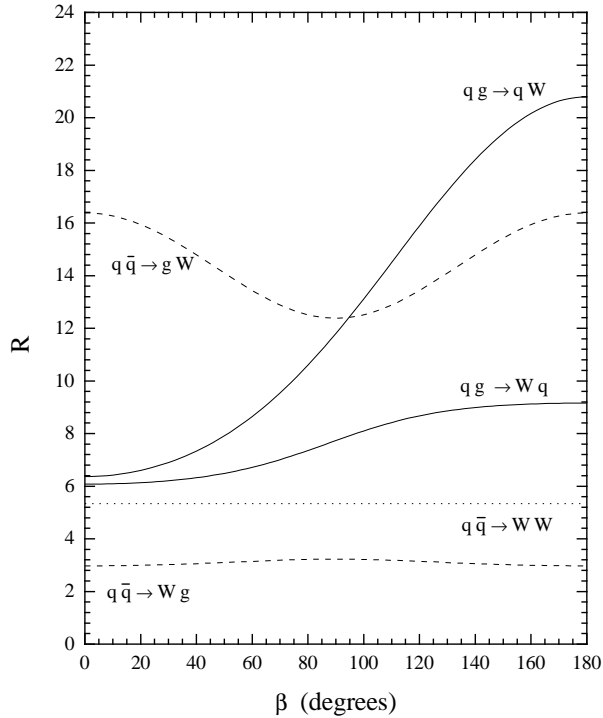,height=12cm}}
\caption{Ratio $\cR$ of the $2\to 3$ and $2 \to 2$ matrix elements
squared  times
$k_T^2/g_s^2$ as a function
of the soft gluon azimuthal angle about the large-$E_T$ jet or $W$, for 
the two subprocesses.}
\label{fig:basic2to2}
\end{center}
\end{figure}

Fig.~\ref{fig:basic2to2} shows the angular ($\beta$) distribution of the
soft gluon jet around each  of the two large-$E_T$ final-state
particles, for $\Delta R = 1$. The quantity plotted is
\begin{equation}
\cR = k_T^2 g_s^{-2}\; {
\overline{\sum} 
\vert\cM_3\vert^2 \over \overline{\sum}\vert \cM_2\vert^2}\; ,
\end{equation}
where the $\vert\cM_n\vert^2$ are the $2\to n$ matrix elements squared
of Eqs.~(\ref{eq:2to3}) and (\ref{eq:2to2}).
 Also shown for reference in Fig.~\ref{fig:basic2to2}  is the
 radiation pattern for the process $q \bar q \to VV$, for which 
 $\cR = 4 C_F = 16/3$, independent of $\beta$. 
 
 For the processes $q \bar q \to Wg $ and $q \bar q \to gW $,
 the distribution is symmetric about $\beta= 90^\circ$ and
 much larger for the latter, as expected. Notice that the soft
gluon distribution around the $W$ in $q \bar q \to W g $ is 
smaller than that around the $W$ in $q \bar q \to WW$; the away-side
gluon jet in the former `drags' the soft gluon away from the $W$
hemisphere.
In the formal limit $\Delta R \to 0$ we find, for this simple kinematic
configuration, 
\begin{eqnarray}
\cR(q \bar q \to W g) & \to  & N_c\left( 1 - \frac{2}{N_c^2} \right) 
= \frac{7}{3}\; ,                        \nonumber \\
\cR(q \bar q \to g W) & \sim   & \frac{4 N_c}{(\Delta R)^2} \; . 
\label{smalldr1}  
\end{eqnarray}
The divergent behaviour  in the latter case reflects the matrix-element
singularity
when the final-state soft and hard gluons are collinear.

The processes $q g \to W q $ and $q g  \to q W $ 
have asymmetric $\beta$ distributions in Fig.~\ref{fig:basic2to2}.
 In both cases the
distribution is larger in the {\it backward} ($\beta > 90^\circ$)
hemisphere, i.e. when the soft gluon moves in the same direction
as the incoming gluon. In fact the distribution for the
$q g \to q W$ case is entirely analogous to that
for $e^+e^- \to q \bar q g$, where the colour flow dynamics
(string \cite{string} or drag \cite{drag} effect)
leads to a suppression of soft radiation between the outgoing
quark and antiquark.

The corresponding $\Delta R \to 0$ limits are (cf. Eq.~(\ref{smalldr1}))
\begin{eqnarray}
\cR(q g \to W q) & \to  & N_c\left( \frac{5}{2} - \frac{1}{2N_c^2} \right) 
= \frac{22}{3}\; ,                        \nonumber \\
\cR(q g \to q W) & \sim   & \frac{4 C_F}{(\Delta R)^2} \; . 
\label{smalldr2}  
\end{eqnarray}

We next address the question of whether the differences between the 
$q \bar q$ and $qg$ subprocesses shown in Fig.~\ref{fig:basic2to2} survive
a more rigorous calculational treatment. We do this in two stages:
first we compare the simple analytic soft-gluon approximation to
the $2\to 3$ matrix elements with the exact result, and then we 
include the effects of phase space, realistic experimental cuts,
parton distributions etc.

\begin{figure}[tb]
\begin{center}
\mbox{\epsfig{figure=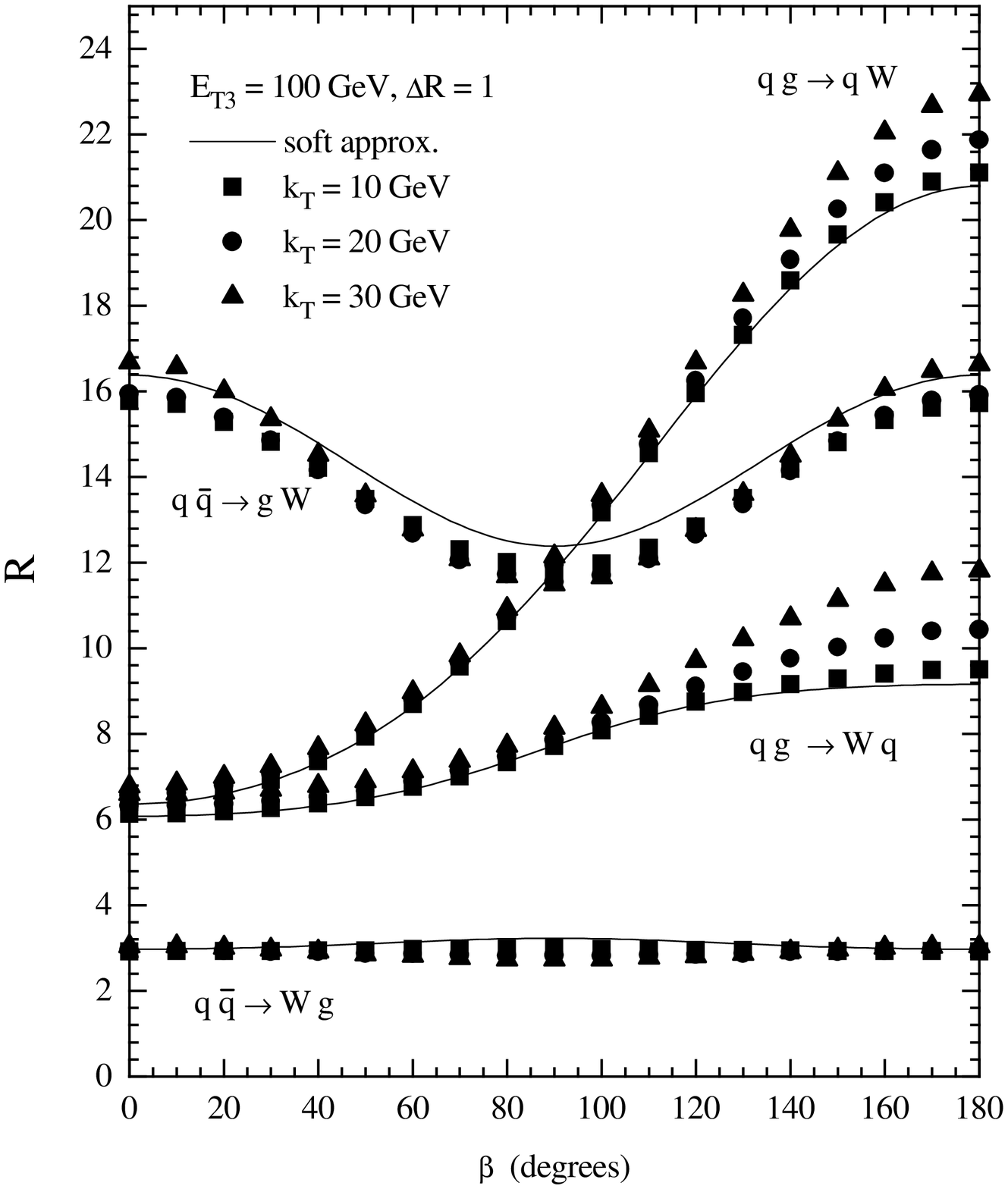,height=12cm}}
\caption{Ratio $\cR$ of the $2\to 3$ and $2 \to 2$ matrix elements 
squared times
$k_T^2/g_s^2$ as a function
of the soft gluon azimuthal angle about the large-$E_T$ jet or W, 
comparing
the exact matrix element result (data points) for various values
of the soft gluon $k_T$ with the $k_T \to 0$ approximations of 
Fig.~\protect\ref{fig:basic2to2} (solid lines).}
\label{fig:exact}
\end{center}
\end{figure}
Fig.~\ref{fig:exact}
compares the exact and approximate calculations of the distributions
of Fig.~\ref{fig:basic2to2}. 
The exact $2\to 3$ matrix elements squared are
calculated   using the spinor techniques of Ref.~\cite{spinor}. 
Exact $2\to 3$ kinematics are used, i.e. having fixed
$p_3$ and $k$ we define $\vec{p_4} = -\vec{p_3} - \vec{k}$ and
$E_4^2 = \vert\vec{p_4}\vert^2 + m_4^2$.   Results
are displayed using three different values of the soft gluon
transverse momentum  $k_T$.
Because the ratio of matrix elements squared is multiplied by $k_T^2$,
the analytic soft-gluon approximation (solid lines) is approached as
$k_T \to 0$. Evidently the soft-gluon approximation gives
a very good representation of the exact result even at quite large
values of $k_T/E_T$. 

The next step is to implement a full cross section calculation, 
including realistic cuts on the final state particles.
We are interested in a final state consisting of a $W$ boson
and two hadronic jets. The $W$ and one jet are produced centrally
with large $E_T$, and the second jet is soft and required to lie
in an annulus in the $(\eta,\phi)$ plane around either the $W$
or the large-$E_T$ jet. The two jets are labelled
according to the ordering of their transverse momenta, $E_T(J) > E_T(j)$.
For illustration, we use the following set of cuts:
\begin{eqnarray}
\label{cuts1}
p_T(W), E_T(J)  >   100\; \GeV \; ,& \quad & 
\vert\eta(W)\vert, \vert\eta(J)\vert < 0.5 \; ,   \nonumber\\
E_T(j)  >  10\; \GeV\; , & \quad & 
0.7 <   R(J \; \mbox{or}\; W,j) < 1.3  \; . 
\label{cuts2}
\end{eqnarray}
In addition, we use the MRS(R2) ($\alpha_s(M_Z^2) = 0.120$)
parton distributions from  Ref.~\cite{MRSR}. One of 
the strong couplings is evaluated at scale $\mu = E_T(J)$
and the other at $\mu = E_T(j)$.
 
\begin{figure}[tb]
\begin{center}
\mbox{\epsfig{figure=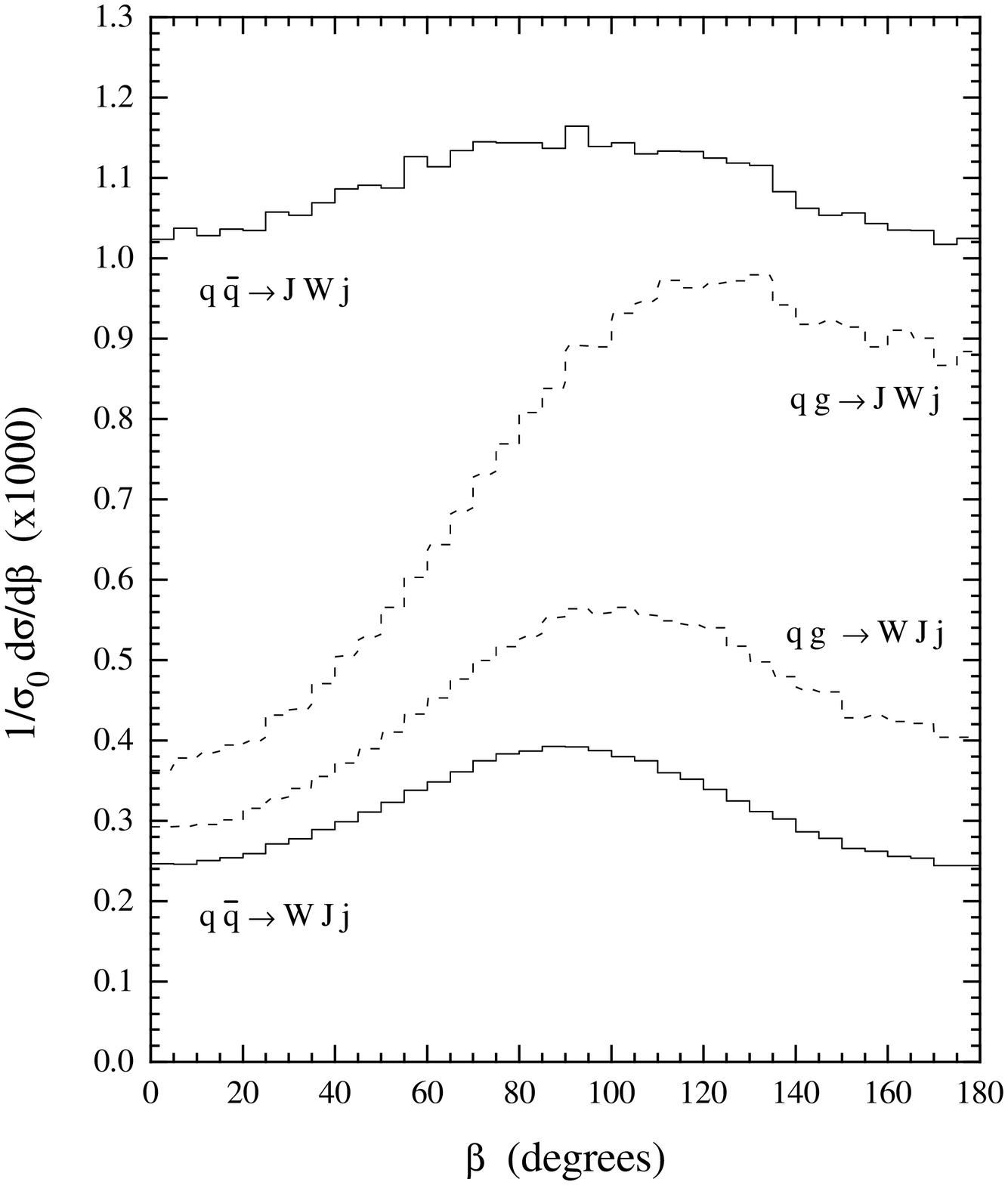,height=12cm}}
\caption{$\beta$ distributions for the $W+2$~jet cross sections
at $\protect\sqrt{s} = 1.8$~TeV, using exact $2\to 3$ matrix elements
with cuts as described in the text,
for different initial states and target final state particles.}
\label{fig:cross1}
\end{center}
\end{figure}
\begin{figure}[tb]
\begin{center}
\mbox{\epsfig{figure=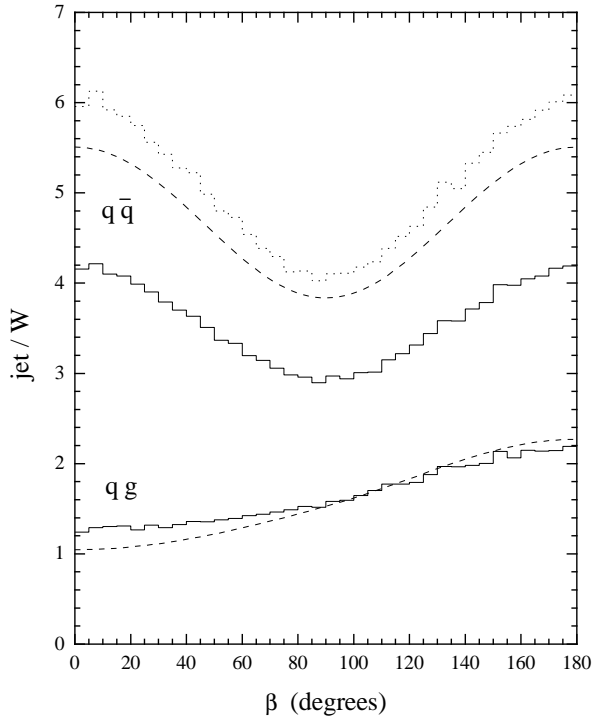,height=12cm}}
\caption{The ratios of jet$/W$ distributions from
Fig.~\protect\ref{fig:cross1} for the $q\bar q$ and $qg$ initial states.
The dashed lines correspond to the soft-gluon approximation, and
the solid histograms correspond to the exact calculation. In the dotted
histogram, the $q \bar q \to W q \bar q$ contribution is absent.}
\label{fig:cross2}
\end{center}
\end{figure}

Fig.~\ref{fig:cross1} shows the $\beta$ distributions calculated
using the above cuts. The separate histograms correspond
to the different initial states ($q \bar q$ and $qg$),
and to the target particle (jet or $W$) at the centre of the annulus.
The curves are normalized to the corresponding $W + 1$~jet lowest-order
cross sections for the corresponding initial state, with the cuts
of (\ref{cuts1}) imposed. It is instructive to compare
Figs.~\ref{fig:cross1} and \ref{fig:exact}. The most noticeable difference
is the suppression of the distributions at $\beta = 0^\circ, 180^\circ$
in the full calculation. This has a simple kinematic explanation.
For fixed values of the rapidities and transverse momenta of the 
$W$ and large-$E_T$ jet, the subprocess energy $\sqrt{\hat s}$ is maximal
when the soft jet rapidity is largest, 
i.e. at $\beta = 0^\circ, 180^\circ$.
These large $\sqrt{\hat s}$ configurations receive an additional
suppression from the parton distribution functions. More importantly,
the {\it relative} size of the various distributions is similar 
to that specified by the soft-gluon antennae  and shown in 
Fig.~\ref{fig:basic2to2}. This is illustrated in Fig.~\ref{fig:cross2},
which shows (solid lines)  the ratio of the `jet' to `$W$' distributions
in Fig.~\ref{fig:cross1} for the $q \bar q $ and $qg$ initial states.

\begin{figure}[tb]
\begin{center}
\mbox{\epsfig{figure=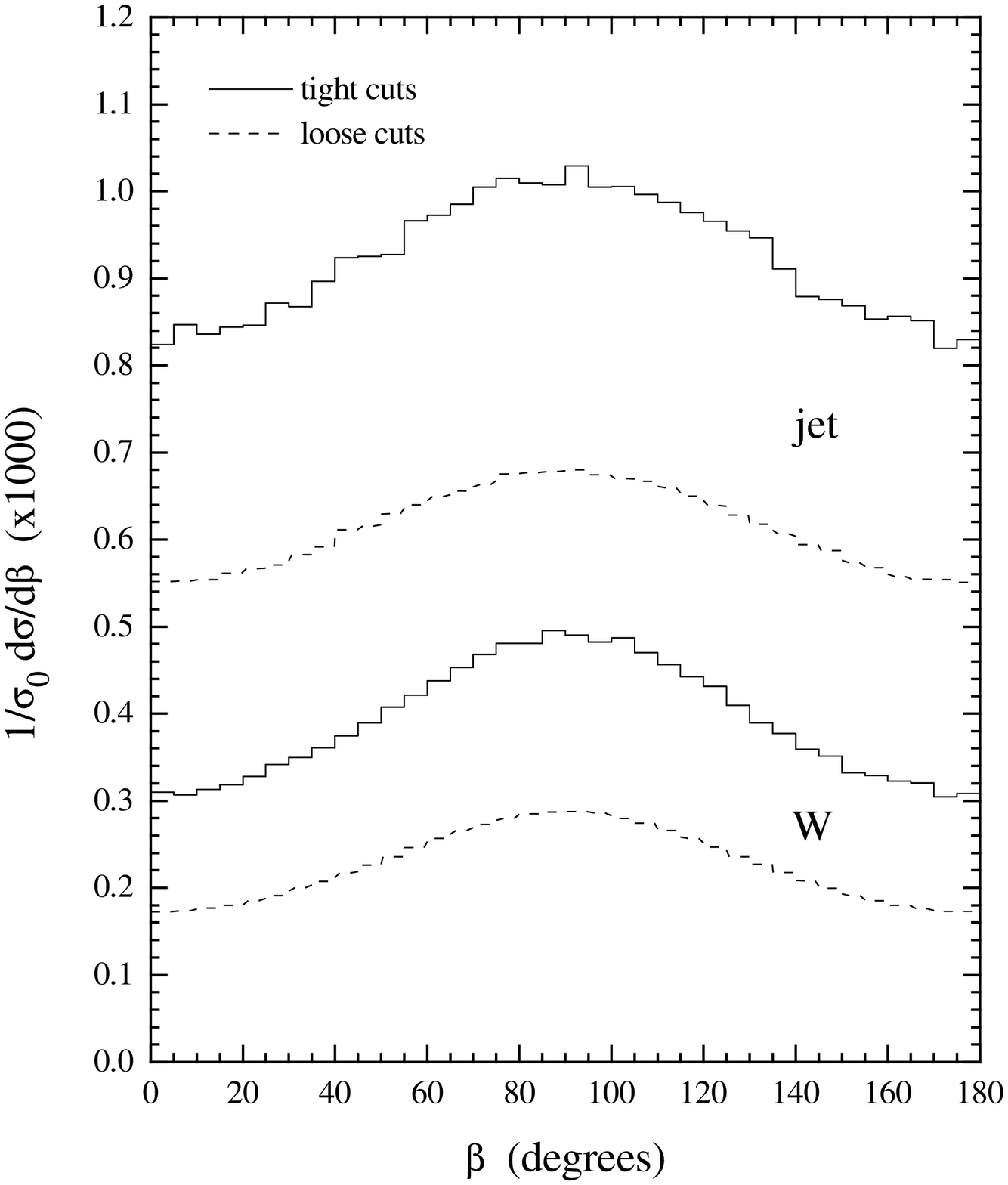,height=12cm}}
\caption{$\beta$ distributions for the $W+2$~jet cross sections
at $\protect\sqrt{s} = 1.8$~TeV, with cuts as described in the text
and including all subprocesses.}
\label{fig:cross3}
\end{center}
\end{figure}

\begin{figure}[tb]
\begin{center}
\mbox{\epsfig{figure=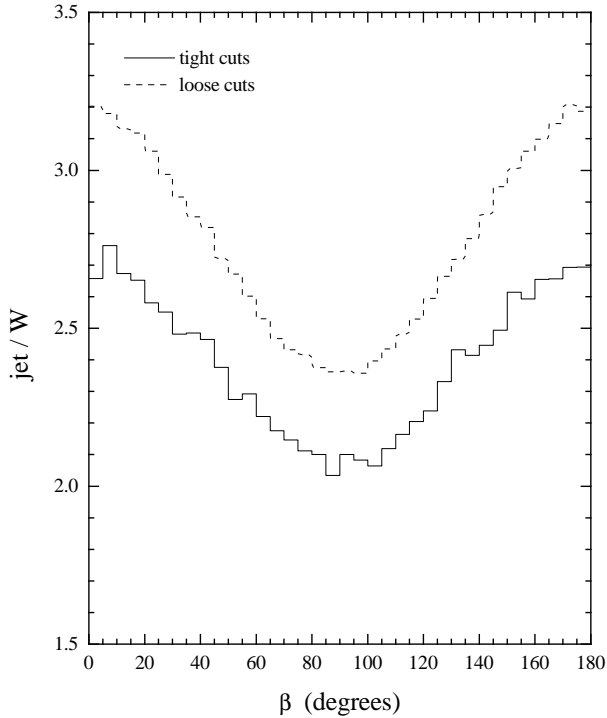,height=12cm}}
\caption{The ratios of jet$/W$ distributions from
Fig.~\protect\ref{fig:cross3} for the two sets
of cuts described in the text.}
\label{fig:cross4}
\end{center}
\end{figure}

For the  $qg$ processes, the agreement between the full calculation
and the soft approximation is excellent.\footnote{Note that in the full
calculation $\Delta R$ is integrated over the range $0.7 < \Delta R
< 1.3$ and the jet and $W$ rapidities are integrated
over the range $\vert\eta\vert < 0.5$, 
whereas in the soft approximation the antennae are calculated
with $\Delta R = 1$ and $\eta = 0$.}
For the $q \bar q$ processes, the soft approximation appears to 
overestimate the result of the full calculation. However, this is due
to the fact that the latter includes contributions from the 
`four-quark' processes $ q \bar q \to W q \bar q$. Although these
are formally subleading in the soft-jet limit ($k_T \to 0$),  they 
make a non-negligible contribution for this choice of parameters and cuts.
Including only the $q \bar q \to W g g $ processes gives instead the dotted
histogram, and agreement with the soft approximation is restored.
Fig.~\ref{fig:cross3} shows the $\beta $ distribution when all subprocesses
are included with their correct parton distribution weighting.\footnote{The
process $g g \to W q \bar q$ is also included but makes a negligible
contribution.} For this choice of cuts, the 
$q \bar q$ and $q g$ contributions are roughly equal, and so the net
distribution interpolates between the separate distributions shown
in Fig.~\ref{fig:cross1}. Note that in Fig.~\ref{fig:cross3} the effect of
including both $qg$ and $gq$ contributions is to symmetrize the
distribution about $\beta = 90^\circ$.
Fig.~\ref{fig:cross4} shows the corresponding prediction for the ratio
of the jet and $W$ distributions, i.e. 
the analogue of Fig.~\ref{fig:cross2}
but including all parton subprocesses and cuts. 

One problem with the analysis described above is that the 
event rate is very small. The requirement of a large $E_T$
central jet and a large $p_T$ central $W$ strongly suppresses
the cross section.
Of course one solution would be to use a less restrictive set
of cuts. Although this means that the asymptotic soft-gluon 
results are less applicable, the qualitative features
of the colour dynamics are still apparent. As an example,  
Figs.~\ref{fig:cross3} and  \ref{fig:cross4} also show
 predictions for a looser set of cuts:
\begin{eqnarray}
\label{cuts3}
p_T(W), E_T(J)  >   30\; \GeV \; ,& \quad & 
\vert\eta(W)\vert, \vert\eta(J)\vert < 1.0 \; ,   \nonumber\\
E_T(j)  >  10\; \GeV\; , & \quad & 
0.7 <   R(J \; \mbox{or}\; W,j) < 1.3  \; . 
\label{cuts4}
\end{eqnarray}
The effect of loosening the cuts is to slightly increase the relative
contribution of the $q\bar q$ subprocess, which has the effect of
 enhancing the jet/$W$ ratio. 
 The following table lists the values of this ratio
for jets produced in the transverse plane ($\beta = \pi/2$)
and in the event plane ($\beta = 0,\pi$).
\begin{center}
\begin{tabular}{|l|c|c|}  \hline
\rule[-1.2ex]{0mm}{4ex}
 & $\beta=  ( 0,\pi )$  & $\beta= \pi/2$ \\ \hline
\rule[-1.2ex]{0mm}{4ex} tight cuts  & 2.7  & 2.1 \\
\rule[-1.2ex]{0mm}{4ex} loose cuts  & 3.2  & 2.4 \\
\hline
\end{tabular}
\end{center}

A higher event rate could also be achieved by studying the distribution
of soft jets in large $E_T$ $\gamma + $jet production, which should exhibit
the same qualitative features as $W+ $jet production, although
there may be experimental problems in obtaining a truly isolated
large $E_T$ photon sample. In fact, an additional theoretical 
advantage of $\gamma + $jet production  
is that there is a stronger dependence
on $E_T$ of the subprocess decomposition, i.e. the relative amounts
of $q \bar q$ and $qg$ scattering. This arises because the average
momentum fraction of the initial-state partons is $\langle x \rangle
\sim 2 E_T/\sqrt{s}$, in contrast to  $\langle x \rangle
\sim (\sqrt{M_W^2 + E_T^2} + E_T)/\sqrt{s}$ for $W+ $jet production.
Note that this argumentation only applies to {\it central} $V+$jet
production. In principle one can also study large $E_T$ events where both 
the $V$ and the jet have large and comparable rapidities, such 
that one of the momentum fractions becomes large and the other small.
The effect of this is to enrich the contribution from $qg$ scattering,
thus modifying the $\beta$  distribution according to Fig.~\ref{fig:cross2}.

An alternative approach, which requires much less luminosity,
is to study the structure of {\it particle} (rather than {\it jet})
flows in the hard scattering events, see 
Refs.~[\ref{ref:DKT}--\ref{ref:book},\ref{ref:drag}].
 Here, unlike in the 
case of jet production, the geometry of the primary parton system
is practically undisturbed, and there is of course no dependence
on the jet-finding algorithm. Using soft particles, the colour coherence
phenomena can  be studied even when the overall statistics 
are limited. However, there is a price to pay. 
The success of the application
of the  analytical perturbative description to the particle
flow distributions is based on the hypothesis of Local Parton Hadron
Duality (LPHD) (see Ref.~\cite{LPHD}), which assumes that the production
of hadrons is governed by the QCD radiophysics of colour flows.
The LPHD hypothesis is very well confirmed by all existing data, thus 
demonstrating that colour coherence effects successfully survive the
hadronization stage, see Ref.~\cite{KO}. 
Care must also be taken when using soft particles to study the colour
coherence properties of the hard scattering to avoid a significant
background contamination from soft particles produced in the
`underlying event'. In practice, this could necessitate a minimum
$p_T$ cut on the registered  hadrons.

\begin{figure}[tb]
\begin{center}
\mbox{\epsfig{figure=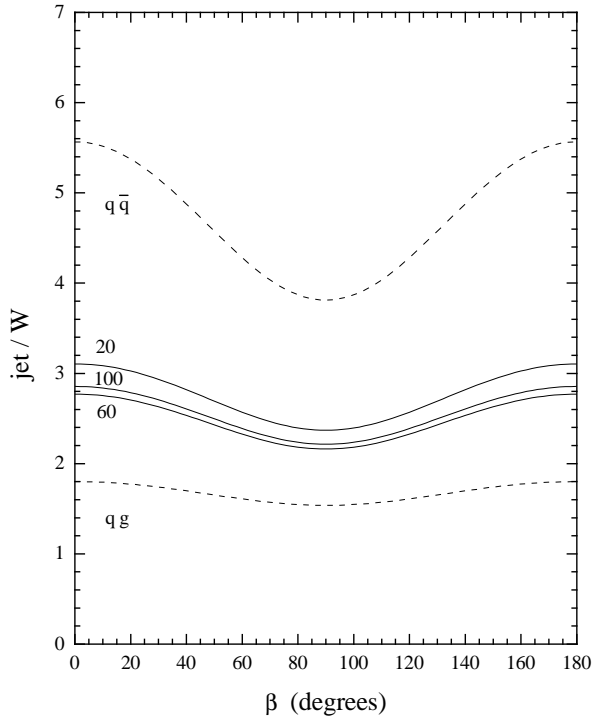,height=12cm}}
\caption{The ratios of jet$/W$ particle distributions,
for different values of the jet $E_T = 20, 60, 100$~GeV. The dashed lines 
are the `pure' $q\bar q$ and $qg$ ratios.}
\label{fig:particles}
\end{center}
\end{figure}

Within the LPHD approach the radiation pattern for wide-angle
soft hadron ($h$) production can be written in the parton-parton
centre-of-mass system as (see Refs.~\cite{DKT,drag})
\begin{equation}
{ d n_h \over d \Omega_{\bf n}} = {E_k^2
\over 8 \pi N_c g_s^{2}}  \; 
{\overline{\sum} \vert\cM_3\vert^2 \over 
\overline{\sum}\vert \cM_2\vert^2}\; 
\left[ N^h_g(E_T)\right]' \; ,
\label{eq:16}
\end{equation}
where
\begin{equation}
\left[ N^h_g(E_T)\right]'  \equiv { d N^h_g(E_T) \over d \ln E_T} 
\end{equation}
and $N^h_g(E_T)$ is the multiplicity of hadrons $h$ in an individual
gluon jet with hardness $E_T$. For the case when there is an experimental
lower limit on the transverse energy of the registered hadron,
\begin{equation}
E_T  \gg  E_T^{\rm min}  \gg \Lambda_{QCD} \; ,
\end{equation}
then the last factor in Eq.~(\ref{eq:16}) becomes
\begin{equation}
\left[ N^h_g(E_T)\right]' \longrightarrow  \left[ N^h_g(E_T)\right]'
-  \left[ N^h_g(E_T^{\rm min})\right]'\; .
\end{equation}
The cascading factor $\left[ N^h_g(E_T)\right]'$
takes into account the fact that the registered hadron $h$ is a part
of the cascade initiated by a soft gluon jet $g$.
It originates in the evolution equation for jet multiplicity
to order $\sqrt{\alpha_s}$ (for details see Refs.~\cite{DKMT,book,drag}):
\begin{equation}
N'_g(E)   =  \int^E {d E_g \over  E_g}
4 N_c {\alpha_s(E_g) \over 2 \pi} N_g(E_g) \; .
\label{evo_formula}
\end{equation}
Note that asymptotically 
\beq
N'_{q}(E)= {C_F \over N_c} \; N'_{g}(E)
\eeq
 with
\beq
{N'_g(E)\over N_g(E)} = \sqrt{{4 N_c \alpha_s(E)\over 2 \pi }}
\; \left[ 1+ O\left( \sqrt{{ \alpha_s\over \pi}}\right) \right]\; .
\label{N_g_formula}
\eeq
For instance, the exact result  for charged particle production is 
(for details see Refs.~\cite{LPHD,DKT-1})
\beq
N^{\rm ch}_g(E) = K^{\rm ch}\; \Gamma(B) \;\left({z\over 2}\right)^{-B+1}
\; I_{B+1}(z) \; ,
\eeq
with 
\bea
z &=& \sqrt{ {16 N_c \over b} Y}\; , \ \ Y = \ln{E\over Q_0}\; , \nonumber \\
B & = &{1\over b}\; \left( \frac{11}{3} N_c + \frac{2n_f}{3N_c^2} \right)
\; \nonumber \\
b & = & {11 N_c - 2 n_f \over 3} \; .
\eea
Here $I_{\nu}$ is the Modified Bessel function.
The values of the parameters $K^{\rm ch} $ and $Q_0$ are obtained from 
fits to the $e^+e^-$ data \cite{DKT-2,KO}:
\beq
K^{\rm ch} \simeq 1.3\; , \ \ Q_0 \simeq 250\; \MeV \; .
\eeq
Finally, the distributions of charged particle flow accompanying
the hard scattering processes $q \bar q \to W g$ and $ q g \to W q$
are given by 
\bea
\left.{ 8\pi d n_h \over 
d \Omega_{\bf n}}\right\vert_{q \bar q \to W g} & = & 
\left( \wh{[14]} + \wh{[24]} -\frac{1}{N_c^2} \wh{[12]} \right)\; 
{d N^{\rm ch}_g \over dY} \; , \nonumber \\
\left.{ 8\pi d n_h \over d \Omega_{\bf n}}\right\vert_{q g \to W q} & = & 
\left( \wh{[12]} + \wh{[24]} -\frac{1}{N_c^2} \wh{[14]} \right)\; 
{d N^{\rm ch}_g \over dY} \; ,
\eea
where 
\beq
\wh{[ij]}  = { 1 - {\bf n}_i \cdot {\bf n}_j \over
 ( 1 - {\bf n} \cdot {\bf n}_i)\;
( 1 - {\bf n} \cdot {\bf n}_j)} \; .
\eeq
Note that these particle distributions take the form of an overall
energy-dependent factor multiplying the {\it same} angular distributions
as in the soft-gluon antenna patterns, Eq.~(\ref{eq:2to3}) and 
Fig.~\ref{fig:basic2to2}.
In studying the relative amounts of hadronic radiation near, say, the
hard jet and the $W$, the ${d N^{\rm ch}_g / dY}$ factors cancel
and the relative angular distribution can be obtained from the curves
in Fig.~\ref{fig:basic2to2}, weighted according to the relative
contributions of the $q\bar q \to W g$ and $q g \to W q$ subprocesses
to the total $W+$jet cross section.
Fig.~\ref{fig:particles} shows the jet$/W$ $\beta$-distribution 
ratio (i.e. the analogue of Fig.~\ref{fig:cross4})
for soft particle production. We assume here that the particles
have much smaller transverse momenta than the hard jet and $W$, so that (i)
there is essentially no phase-space suppression of the distributions,
and (ii) the $q\bar q \to  W q \bar q$ process makes a negligible
contribution. In addition, the distribution is integrated
over the annuli\footnote{Recall that the curves
in Fig.~\ref{fig:basic2to2} are evaluated at $\Delta R = 1$.}
 $0.7 < \Delta R < 1.3$ around the jet and the $W$,
which are produced centrally,
$\vert\eta\vert < 0.5$. The  curves correspond to different values
of the hard jet $E_T$. In fact for  $E_T$ values in the
range $20\; \GeV \lapprox E_T \lapprox 100\; \GeV$ the $q\bar q$ and
$qg$ subprocesses give approximately equal contributions.
The dashed lines in Fig.~\ref{fig:particles}
are the limiting `pure' $q\bar q$ and $qg$ ratios. Notice that
the particle jet$/W$ ratios are very similar in magnitude to the 
soft-jet ratios shown in Fig.~\ref{fig:cross4}. i.e. when expressed in
terms of this ratio, the colour coherence
effects appear to be universal. However, there is an important
caveat to this conclusion. In this study we are only calculating
the {\it perturbative} contribution to the particle and jet flow, i.e.
the contribution associated with gluon emission 
from the single hard scatter
which produces the $W$ and the jet. In practice, there will also be a 
(non-perturbative) contribution from the `underlying event'.
The lower the particle/jet $k_T$ threshold, the more important 
this contribution is likely to be. We would further expect
this background contribution to be approximately uniform in the 
$(\eta,\phi)$ plane, thereby increasing the $\beta$ distributions
by a constant amount. This would have the effect of {\it decreasing}
the jet$/W$ ratios shown in Figs.~\ref{fig:cross4} and \ref{fig:particles}.
Since we are assuming that in practice the transverse momentum
threshold for particle production
will be much lower than for soft jet production, we would expect to see
a larger effect in the former jet$/W$ ratio.

In conclusion, the aim of this paper has been
to exemplify the application of the radiophysics
of colour flows to the partonometry of  $V(=\gamma, W,Z) +$jet
events in hadron-hadron collisions.
As was appreciated long ago \cite{DKT,DKMT},
the colour interference effects accompanying such processes
can be especially spectacular.
They may well play for hadron colliders the same role as the celebrated
string \cite{string} and drag \cite{drag} phenomena observed
in three jet production in $e^+e^-$ annihilation.
Clear experimental observation of the effects discussed
in this paper, and the quantitative agreement with the basic formulae
presented here, would lend strong support to the idea that
hadronic antenna patterns  can provide a valuable diagnostic tool
for elucidating the underlying dynamics in multi-jet high $E_T$
events. The first experimental results on $W+$jet production from
the D0 collaboration \cite{D0dpf96,D0Melanson}  look very promising.
 
\vspace{0.5cm}
\noindent {\bf Acknowledgements} \\
We thank J.~Ellis, A.~Goshaw and  N.~Varelas
for useful comments and discussions. This work  was
supported in part by the UK PPARC and the EU Programme
``Human Capital and Mobility'', Network ``Physics at High Energy
Colliders'', contract CHRX-CT93-0357 (DG 12 COMA).

\end{document}